\documentclass[authoryear,preprint,review,12pt]{elsarticle}
\usepackage[margin=1in]{geometry}
\usepackage{amsmath}
\usepackage[english]{babel}
\usepackage{graphicx}
\usepackage{caption}
\usepackage{array}
\usepackage{rotating}
\usepackage{color}
\usepackage{setspace}
\usepackage{multirow}
\usepackage[caption=false]{subfig}

\usepackage{lipsum}
\usepackage{amsfonts}
\usepackage{epstopdf}
\usepackage{float}
\newtheorem{theorem}{Theorem}

\renewcommand{\vec}[1]{\mathbf{#1}}
\newcommand{\svec}[1]{\boldsymbol{#1}}
\newcommand{\Yv}{\vec{Y}}

\newcommand{\tauv}{\svec{\tau}}
\newcommand{\gammav}{\svec{\gamma}}

\newcommand{\betav}{\svec{\beta}} 

\newcommand{\thetav}{\svec{\theta}}

\begin{document}
\begin{center}{\Huge Bayesian Crossover Designs for Generalized Linear Models}\end{center}
\vspace{.5em}

\begin{center}{{\Large } Satya Prakash Singh, Siuli Mukhopadhyay{\footnote {Corresponding author. Email: siuli@math.iitb.ac.in}}\\
{\it Department of Mathematics, Indian Institute of Technology Bombay,}\\ {\it Mumbai 400 076, India}}\end{center}
\vspace{.25em}

\date{}

\doublespacing

\hrule

\vspace{.5em}

\noindent {\bf Abstract}

This article discusses  optimal Bayesian crossover designs for generalized linear models. Crossover trials with $t$ treatments and $p$ periods, for $t<=p$, are considered. The designs proposed in this paper minimize the log determinant of the variance of the estimated treatment effects over all possible allocation of the $n$ subjects to the treatment sequences. It is assumed that the $p$ observations from each subject are mutually correlated while the observations from different subjects are uncorrelated. Since main interest is in estimating the treatment effects, the subject effect is assumed to be nuisance, and generalized estimating equations are used to estimate the marginal means. To address the issue of parameter dependence a Bayesian approach is employed. Prior distributions are assumed on the model parameters which are then incorporated into the $D_A$-optimal design criterion by integrating it over the prior distribution. Three case studies, one with binary outcomes in a $4\times4$ crossover trial, second one based on count data for a $2\times2$ trial and a third one with Gamma responses in a $3\times2$ crossover trial are used to illustrate the proposed method. The effect of the choice of prior distributions on the designs is also studied. A general equivalence theorem
is stated to verify the optimality of designs obtained.
\vspace{.5em}

\vspace{.5em}

\vspace{.5em}
\hrule

\vspace{1.5em}
\section{Introduction}
In this article we introduce Bayesian optimal crossover designs for generalized linear models (GLMs). Crossover trials with $t$ treatments and $p$ periods, for $t\leq p$ are considered. The designs selected minimize the log determinant of the variance-covariance matrix of the treatment effects, over all possible allocation of the $n$ subjects to the treatment sequences.  Due to the dependence of the variance matrix on the model parameters a Bayesian approach is proposed.

Crossover designs were originally developed to be used in agricultural sciences (\cite{Cochran1939}). Later, these repeated measurement designs were found to be useful in many other fields, such as  pharmaceutical and clinical trials,  bioequivalence and biological studies. Optimal crossover  designs for normal response have been studied by many reserachers, namely \cite{Hedayat1975,Hedayat1978}, \cite{cheng1980}, \cite{Laska1983}, \cite{Laska1985}, \cite{Stufken1991}, \cite{Carrire1993}, \cite{Kushner1997, Kushner1998} and \cite{Carriere_2000}. For a detailed review of crossover designs, we would like to refer to the paper by \cite{Bose_2013} and books by \cite{Bose2009}, \cite{Senn_2002} and \cite{Jones_2014}.

Most of the available literature on optimal  crossover designs (as discussed above)  mainly focuses on normal responses.
However, in  biological studies, very often we find responses that are non-normal (\cite{Layard_1978} and \cite{Forster1992}) and have to be modeled using a generalized linear model  (GLM). While methods for analyzing GLM data arising from crossover trials are available in  \cite{Senn_2002} and \cite{Jones_2014}, the question of designing such studies for GLMs in an optimal manner does not seem to have been much explored in the statistical literature. \cite{Waterhouse_2006} studied optimal $2\times 2$ crossover trial for  binary data in some special cases, like the carryover effect is proportional to the direct treatment effect and no period effects are considered. Adaptive crossover designs restricted to two period two treatment binary data useful in clinical trials have also been investigated by \cite{Bandyopadhyay2009}.



In this article, we study optimal Bayesian crossover designs  for GLMs.  Three case studies based on non-normal responses are used to illustrate the proposed methodology. Generalized estimating equations of Liang and Zeger (1986) are used to estimate the marginal means. The correlation between observations within subjects are modeled using a ``working correlation structure", which is  assumed to be compound symmetric or auto regressive in nature. Since the main interest is in estimating the treatment effects, the subject effects as taken as  nuisance parameters. As in all GLM designs, the variance of the treatment effect estimator depends on the model parameters. To address the  issue of the parameter dependence and obtain robust designs we propose
the Bayesian approach to design selection. Bayesian designs have been a popular choice whenever the variance-covariance matrix depends on the model parameters, for some references see  (\cite{Chaloner1989}, \cite{Dette1994}, \cite{Woods2011} and \cite{Mylona2014} ). In our approach, a prior distribution is assumed on the model parameters, which is then incorporated into an appropriate objective function
(variance of the treatment contrast) by integrating and averaging over the prior distribution.
Similar to our Bayesian design criterion, an average criterion called $A$-criterion have been used before for  crossover designs for normal responses by (\cite{Kempton2001}, \cite{Baily2006}, \cite{Zheng2013} and \cite{Li2015}).

\section{Case studies}
For illustration purpose we consider three case studies based on crossover trials  involving binary, count and  Gamma responses.
\subsection{A four periods four treatments binary response crossover trial}
The first case study presented here is from a trial based on the four-period, four treatment
Williams design. It has been reported in  \cite{Kenward1992}. The four treatments  are denoted by $A, B, C$ and $D$. Eighty subjects are randomly assigned to the four treatment sequences \{$ABCD, BDAC, CADB, DCBA$\}, with about twenty subjects allocated to each treatment sequence. The response is a binary outcome taking values 1 and 0 based on patient relief and no relief, respectively.

The research question which arises from the above case study is why did the experimenter select the 4 treatment sequences \{$ABCD, BDAC, CADB, DCBA$\} forming a Williams design (\cite{Williams1949}).  Is this the best possible selection of treatment sequences? The book by \cite{Bose2009}, page 40 shows that for normal response crossover models, for the 4 treatment and 4 periods case, Williams design is the optimal design.  But can we be sure that the same design applies to a binary response crossover framework as well? Does the selected design change if the correlation structure between observations change say, from equicorrelated to auto regressive structure?

\subsection{Two periods two treatments Poisson response crossover trial}
This study is based on an example described in \cite{Layard_1978}. Two  drugs, standard drug A and an innovation  drug B, is administered for controlling angina in 20 patients. It is known that the innovative drug $B$ is no worse than the standard drug $A$. For a given patient, number of angina attacks on weekly basis is assumed to follow a Poisson distribution (\cite{Layard_1978}). Number of attacks for each patient of consecutive two weeks are recorded. Treatment sequences considered are \{$AB$, $BA$\} and 10 patients are assigned to each of the treatment sequences. This is a 2-treatments 2-periods crossover trial.

As in case study I, the question arises why does the experimenter choose the design $\{AB, BA$\}. Is this the best or most efficient design under the repeated measures setup when responses follow a Poisson distribution?

\subsection{Three periods two treatments Gamma response trial}
The length of hospital stay is an important measure of the success of hospital activity, costs incurred by patients and the treatment administered to a patient. However, its empirical distribution is often right skewed and a Gamma distribution with a log link has been seen to be a good fit (\cite{Faddy2009}). In this case study we consider a crossover trial where two treatments are applied over three periods and length of hospital stay, assumed to having a Gamma distribution, is the primary end point.

As in the earlier two case studies, we investigate the best design for a two treatment three periods design with a gamma response.

\section{The model}

We consider experiments where there are $t$ treatments and $n$ subjects, and $p$ repeated measurements are taken from each subject.  The observations from each subject may be correlated. The marginal distribution of the response $Y_{ij}$  is described by a working generalized linear model with the following three components (\cite{Liang1986}):
\begin{enumerate}
\item  $Y_{ij}$ has a distribution from the exponential family form,
\begin{equation}
  f(y_{ij}|\phi_{ij},\psi)  = \mbox{exp}\left\{ \left[y_{ij}\phi_{ij} - b(\phi_{ij}) + c(y_{ij})\right]\psi + d(y_{ij}, \psi) \right\}\label{efd1}
\end{equation}
where $\phi_{ij}$ is a function of the model parameters, $b(\cdot)$, $c(\cdot)$ and $d(\cdot)$ are known functions and $\psi$ is the dispersion parameter.
It can be shown that:
${E}(Y_{ij})=\mu_{ij} = \frac{db(\phi_{ij})}{d\phi_{ij}}$ and
${Var}(Y_{ij})=\frac{d^2 b(\phi_{ij})}{d\phi^{2}_{ij}}/\psi$.
\item The linear predictor $\eta_{ij}$ in a repeated measures setup can be written as (\cite{Bose2009}),
\begin{equation}\label{margmodel}
\eta_{ij}=\nu+\beta_i+\tau_{d(i,j)}+\gamma_{d(i-1,j)};\,i=1\,\ldots,p,\,j=1,\ldots,n,
\end{equation}
where
$\nu$ is the fixed unknown parameter, $\beta_i$ represents the effect of the $i$th period, $\tau_s$ is the direct effect due to treatment $s$ and $\gamma_s$ is the carryover effect due to treatment $s$,  $s=1,\ldots ,t$. It is assumed that $\gamma_{d(0,j)}=0$.
\item The mean of $y_{ij}$ denoted by $\mu_{ij}$ is related to $\eta_{ij}$ through a link function $g$, where $g(\mu_{ij})=\eta_{ij}$ and the inverse of $g$ exists.
\end{enumerate}

\subsection{Estimation}
 Regression coefficients as well as their variances are estimated by the GEE approach of \cite{Liang1986} and \cite{Zeger1988}.
Due to observations from the same subject being correlated, a ``working correlation" matrix, $R(\alpha)$, is used to describe the dependencies between repeated observations from a subject. Here $\alpha$ is a vector of length $l$, which fully characterizes $R(\alpha)$ (\cite{Liang1986}). For cases where $R(\alpha)$ is the true correlation matrix of $\Yv_j = (Y_{1j}, \cdots, Y_{pj})'$, the covariance of $\Yv_j$ is
\begin{equation}\label{varmatrix}
V_j=A_j^{1/2}R(\alpha)A_j^{1/2},
\end{equation}
$A_j=\text{diag}(\text{Var}(Y_{1j}),\ldots,\text{Var}(Y_{pj}))$.  If the correlation structure is compound symmetric that is corr($Y_{ij}, Y_{i'j}$) = $\alpha$ for all $i\neq i'$, then $l = 1$, if the correlation structure is left unspecified then $l = \frac{p(p-1)}{2}$.
Also, the asymptotic variance for the GEE estimator $\hat{\thetav}$ (see \cite{Zeger1988}, equation (3.2)) is
\begin{equation}\label{fvarmatrix}
Var(\hat{\thetav})=
\left[\sum_{j=1}^{n}\frac{\partial\mu'_j}{\partial\thetav} V_{j}^{-1}\frac{\partial\mu_j}{\partial\thetav}\right]^{-1},
\end{equation}
where $\thetav = (\nu, \betav', \tauv', \gammav')'$, $\betav' = (\beta_{1},\cdots\beta_{p})$, $\tauv' = (\tau_{1}, \cdots, \tau_{t})$ and $\gammav' = (\gamma_{1}, \cdots, \gamma_{t})$.

 However, if the true correlation structure varies from the ``working correlation" structure, then $Var(\hat{\thetav})$ is given by the sandwich formula (\cite{Zeger1988}, equation (3.2))
\begin{equation}\label{WTmatrix}
Var(\hat{\thetav})=
\left[\sum_{j=1}^{n}\frac{\partial\mu'_j}{\partial\thetav} V_{j}^{-1}\frac{\partial\mu_j}{\partial\thetav}\right]^{-1}\left[\sum_{j=1}^{n}\frac{\partial\mu'_j}{\partial\thetav} V_{j}^{-1}Cov(Y_j)V_{j}^{-1}\frac{\partial\mu_j}{\partial\thetav}\right]\left[\sum_{j=1}^{n}\frac{\partial\mu'_j}{\partial\thetav} V_{j}^{-1}\frac{\partial\mu_j}{\partial\thetav}\right]^{-1}.
\end{equation}

For the crossover model (1), the $i$th element of $\frac{\partial\mu_j}{\partial\thetav}$  is $\frac{\partial\mu_{ij}}{\partial\thetav}=x'_{ij} \frac{\partial g^{-1}(\eta_{ij})}{\partial \eta_{ij}}$, where $x'_{ij}$ is the $i$th row of  $X_j$ for $i=1,\ldots,p$. The design matrix is $X_j =[1_{p} \;P_j \;T_j\;F_j]$, where
$P_j=I_p$; $T=(T'_1,\ldots,T'_n)'$, where $T_j$ is a $p\times t$ matrix with its $(i,s)$th entry equal to 1 if subject $j$ receives the direct effect of the treatment $s$ in the $i$th period and zero otherwise; $F=(F'_1,\ldots,F'_n)'$, where $F_j$ is a $p\times t$ matrix with its $(i,s)$th entry equal to 1 if subject $j$ receives the carryover effect of the treatment $s$ in the $i$th period and zero otherwise.

\subsection{Specific case: Bernoulli distribution}
If $Y_{ij}\sim \mbox{Bernoulli}(\mu_{ij})$, then the probability mass function of  $Y_{ij}$ is:
\begin{equation*}
f(y_{ij}|\mu_{ij}) =  \mbox{exp}\left\{y_{ij}\mbox{log}\frac{\mu_{ij}}{1-\mu_{ij}}+\mbox{log}(1-\mu_{ij})\right\}
\end{equation*}

Comparing with equation (\ref{efd1}), we get $\phi_{ij} =  \mbox{log}\frac{\mu_{ij}}{1-\mu_{ij}}$, $b(\phi_{ij}) = -\mbox{log}(1-\mu_{ij}) = \mbox{log}(1+\mbox{exp}(\phi_{ij}))$, $c(y_{ij}) = 0$, $\psi = 1$ and $d(y_{ij}, \psi) = 0$. The mean of $Y_{ij}$ is $\mbox{E}(Y_{ij}) = \mu_{ij} = \frac{db(\phi_{ij})}{d\phi_{ij}} = \frac{\mbox{exp}(\phi_{ij})}{1+\mbox{exp}(\phi_{ij})}$,
 and $\mbox{Var}(Y_{ij}) = \frac{d^2 b(\phi_{ij})}{d\phi^{2}_{ij}}/\psi = \frac{\mbox{exp}(\phi_{ij})}{(1+\mbox{exp}(\phi_{ij}))^{2}} = \mu_{ij}(1-\mu_{ij})$.

Considering the logit link function to relate the linear predictor $\eta_{ij}$ to the mean $\mu_{ij}$, $g(\mu_{ij}) = \mbox{log}\frac{\mu_{ij}}{1-\mu_{ij}}$. Thus $g^{-1}(\eta_{ij}) = \frac{\mbox{e}^{\eta_{ij}}}{1+\mbox{e}^{\eta_{ij}}}$, and the $i$th component of $\frac{\partial\mu_j}{\partial\thetav}$  is $\frac{\partial\mu_{ij}}{\partial\thetav}=x'_{ij} \frac{\partial g^{-1}(\eta_{ij})}{\partial \eta_{ij}} = x'_{ij}\frac{\mbox{e}^{\eta_{ij}}}{(1+\mbox{e}^{\eta_{ij}})^{2}} = x'_{ij}\mu_{ij}(1-\mu_{ij})$. This implies $\frac{\partial\mu_j}{\partial\thetav} = D_{j}X_{j}$, where $D_{j}$ is the diagonal $p\times p$ matrix with  elements $\mu_{ij}(1-\mu_{ij}), i = 1,\cdots, p$. The matrix $A_j$ defined in equation (\ref{varmatrix}) is same as $D_{j}$ in this case. Using equation (\ref{fvarmatrix}), the asymptotic information matrix is:

\begin{eqnarray*}
\sum_{j=1}^{n}\frac{\partial\mu'_j}{\partial\thetav} V_{j}^{-1}\frac{\partial\mu_j}{\partial\thetav} &= & \sum_{j=1}^{n}X'_{j}D_{j}A_{j}^{-1/2}R^{-1}(\alpha)A_{j}^{-1/2}D_{j}X_{j}\\
& = & \sum_{j=1}^{n}X'_{j}A_{j}^{1/2}R^{-1}(\alpha)A_{j}^{1/2}X_{j}.
\end{eqnarray*}

\subsection{Specific case: Poisson distribution}
If $Y_{ij}\sim \mbox{Poisson}(\mu_{ij})$, then the probability mass function of $Y_{ij}$ is:
\begin{equation*}
f(y_{ij}|\mu_{ij}) = \mbox{exp}\left\{y_{ij}\mbox{log}(\mu_{ij}) - \mu_{ij} - \mbox{log}(y_{ij}!)\right\}
\end{equation*}

Comparing with equation (\ref{efd1}), we get $\phi_{ij} = \mbox{log}(\mu_{ij})$, $b(\phi_{ij}) = \mu_{ij} = \mbox{e}^{\phi_{ij}}$, $c(y_{ij}) = -\mbox{log}(y_{ij}!)$, $\psi = 1$ and $d(y_{ij}, \psi) = 0$. The mean of $Y_{ij}$ is $\mbox{E}(Y_{ij})$ = $\mu_{ij} = \mbox{e}^{\phi_{ij}}$ and  $\mbox{Var}(Y_{ij}) = \mbox{e}^{\phi_{ij}} = \mu_{ij}$.

Using the log link we obtain, $g(\mu_{ij}) = \mbox{log}(\mu_{ij}) = \eta_{ij}$, and the $i$th component of $\frac{\partial\mu_j}{\partial\thetav}$  is $\frac{\partial\mu_{ij}}{\partial\thetav}=x'_{ij} \frac{\partial g^{-1}(\eta_{ij})}{\partial \eta_{ij}} = x'_{ij}\mbox{e}^{\eta_{ij}} = x'_{ij}\mu_{ij}$. This implies $\frac{\partial\mu_j}{\partial\thetav} = D_{j}X_{j}$, where $D_{j}$ is the diagonal $p\times p$ matrix with  elements $\mu_{ij} , i = 1,\cdots, p$. The matrix $A_j$ defined in equation (\ref{varmatrix}) is again same as $D_{j}$ in this case. The asymptotic information matrix is:

\begin{eqnarray*}
\sum_{j=1}^{n}\frac{\partial\mu'_j}{\partial\theta} V_{j}^{-1}\frac{\partial\mu_j}{\partial\theta} &= & \sum_{j=1}^{n}X'_{j}D_{j}A_{j}^{-1/2}R^{-1}(\alpha)A_{j}^{-1/2}D_{j}X_{j}\\
& = & \sum_{j=1}^{n}X'_{j}A_{j}^{1/2}R^{-1}(\alpha)A_{j}^{1/2}X_{j}.
\end{eqnarray*}

\subsection{Specific case: Gamma distribution} \label{gamma}
If $Y_{ij}\sim \mbox{Gamma}(\kappa, \lambda_{ij})$, where $\kappa>0$ is the shape parameter and $\lambda_{ij}>0$ is the rate parameter. Then the probability density function of $Y_{ij}$  is:
\begin{equation*}
f(y_{ij}|\lambda_{ij},\kappa) = \mbox{exp}\left\{\left[y_{ij}\left(-\frac{\lambda_{ij}}{\kappa}\right) + \mbox{log}\left(\frac{\lambda_{ij}}{\kappa}\right) + \mbox{log}(y_{ij})\right]{\kappa} + \kappa\mbox{log}(\kappa)- \mbox{log}(y_{ij}) -\mbox{log}\Gamma \kappa\right\}
\end{equation*}
Comparing with equation (\ref{efd1}), we get $\phi_{ij} = -\frac{\lambda_{ij}}{\kappa}$, $b(\phi_{ij}) = -\mbox{log}(\frac{\lambda_{ij}}{\kappa}) = -\mbox{log}(-\phi_{ij})$, $c(y_{ij}) = \mbox{log}(y_{ij})$, $\psi = \kappa$ and  $d(y_{ij}, \psi) = \kappa\mbox{log}(\kappa)-\mbox{log}{(y_{ij})}({\Gamma \kappa}) = \psi\mbox{log}(\psi)-\mbox{log}{(y_{ij})}({\Gamma \psi})$. The mean of $Y_{ij}$ is  $\mbox{E}(Y_{ij}) =  \mu_{ij} = \kappa/\lambda_{ij}$ and $\mbox{Var}(Y_{ij}) = \frac{k}{\lambda_{ij}^{2}} = \frac{\mu_{ij}^2}{\kappa}$.

In case of  a log link function, $g(\mu_{ij}) = \mbox{log}(\mu_{ij}) = \eta_{ij}$.
The $i$th component of $\frac{\partial\mu_j}{\partial\thetav}$  is $\frac{\partial\mu_{ij}}{\partial\thetav}=x'_{ij} \frac{\partial g^{-1}(\eta_{ij})}{\partial \eta_{ij}} = x'_{ij}\mbox{e}^{\eta_{ij}} = x'_{ij}\mu_{ij}$. This implies $\frac{\partial\mu_j}{\partial\thetav} = D_{j}X_{j}$, where $D_{j}$ is the diagonal $p\times p$ matrix with elements $\mu_{ij}, i = 1,\cdots, p$. The matrix $A_j$ defined in equation (\ref{varmatrix}) is  diagonal $p \times p$ matrix with elements $\frac{\mu_{ij}^2}{\kappa}, i = 1,\cdots, p$. The asymptotic information matrix is:

\begin{eqnarray*}\label{normallikematrix}
\sum_{j=1}^{n}\frac{\partial\mu'_j}{\partial\theta} V_{j}^{-1}\frac{\partial\mu_j}{\partial\theta} &= & \sum_{j=1}^{n}X'_{j}D_{j}A_{j}^{-1/2}R^{-1}(\alpha)A_{j}^{-1/2}D_{j}X_{j} \nonumber\\
& = & \sum_{j=1}^{n}X'_{j}\left\{\sqrt{\kappa}I_{p}\right\}R^{-1}(\alpha)\left\{\sqrt{\kappa}I_{p}\right\}X_{j} \nonumber\\
& = & \kappa \sum_{j=1}^{n}X'_{j}R^{-1}(\alpha)X_{j}.
\end{eqnarray*}

In case of  a reciprocal link function, $g(\mu_{ij})  = \frac{1}{\eta_{ij}}$.
The $i$th component of $\frac{\partial\mu_j}{\partial\thetav}$  is $\frac{\partial\mu_{ij}}{\partial\thetav}=x'_{ij} \frac{\partial g^{-1}(\eta_{ij})}{\partial \eta_{ij}} = -\frac{x'_{ij}}{{\eta_{ij}^2}} = -x'_{ij}\mu_{ij}^2$. This implies $\frac{\partial\mu_j}{\partial\thetav} = D_{j}X_{j}$, where $D_{j}$ is the diagonal $p\times p$ matrix with  elements $-\mu_{ij}^2, i = 1,\cdots, p$. The matrix $A_j$ defined in equation (\ref{varmatrix}) is A diagonal $p \times p$  matrix with elements $\frac{\mu_{ij}^2}{\kappa}, i = 1,\cdots, p$. The asymptotic information matrix can be written as:

\begin{eqnarray*}
\sum_{j=1}^{n}\frac{\partial\mu'_j}{\partial\theta} V_{j}^{-1}\frac{\partial\mu_j}{\partial\theta} &= & \sum_{j=1}^{n}X'_{j}D_{j}A_{j}^{-1/2}R^{-1}(\alpha)A_{j}^{-1/2}D_{j}X_{j}\\
& = & \sum_{j=1}^{n}X'_{j}\left\{-\sqrt{\kappa}D_{j}^{*}\right\}R^{-1}(\alpha)\left\{-\sqrt{\kappa}D_{j}^{*}\right\}X_{j}\\
& = & \kappa \sum_{j=1}^{n}X'_{j}D_{j}^{*}R^{-1}(\alpha)D_{j}^{*}X_{j},
\end{eqnarray*}
where $D_{j}^*$ is the diagonal matrix with diagonal elements $\mu_{ij}, i = 1,\cdots, p$.

Note that the shape parameter $\kappa$ is a multiplicative constant in the expression of the information matrices and hence does not affect design selection.
\section{Approximate designs}
For finding optimal crossover designs for the logistic model we use the approximate theory as in \cite{Laska1983} and \cite{Kushner1997, Kushner1998}.
Suppose ${\Omega}$ is the set of treatment sequences of the form $\omega=(t_1,\ldots,t_p)', \,t_i\in \{1,\ldots,t\}$, and
 $n_\omega$ is the number of subjects assigned to sequence $\omega$. Then, $n=\sum_{\omega\in\Omega}n_\omega,n_\omega\geq 0$. A design $\zeta$ in approximate theory is specified by the set $\{p_\omega,\omega\in\Omega\}$ where $p_\omega=n_\omega/n$, is the proportion of subjects assigned to treatment sequence $\omega$.

The matrices $T_j$ and $F_j$ depend only on the treatment sequence $\omega$ to which the $j$th subject is assigned,  so $T_j=T_\omega,\,F_j=F_\omega$, implying, $X_j=X_\omega$. Thus, the variance of $\hat{\thetav}$ is
\begin{equation}\label{optcrt}
Var_\zeta(\hat{\thetav})=\left[\sum_{\omega\in\Omega}np_\omega\frac{\partial\mu'_\omega}{\partial\thetav} V_{\omega}^{-1}\frac{\partial\mu_\omega}{\partial\thetav}\right]^{-1}\left[\sum_{\omega\in\Omega}np_\omega\frac{\partial\mu'_\omega}{\partial\thetav} V_{\omega}^{-1}Cov(Y_\omega)V_{\omega}^{-1}\frac{\partial\mu_\omega}{\partial\thetav}\right]\left[\sum_{\omega\in\Omega}np_\omega\frac{\partial\mu'_\omega}{\partial\thetav} V_{\omega}^{-1}\frac{\partial\mu_\omega}{\partial\thetav}\right]^{-1}.
\end{equation}
If the true correlation of $Y_j$ is equal to $R(\alpha)$ then we have a much simpler form,
\begin{equation}\label{infmat}
Var_\zeta(\hat{\thetav})=\left[\sum_{\omega\in\Omega}np_\omega\frac{\partial\mu'_\omega}{\partial\thetav} V_{\omega}^{-1}\frac{\partial\mu_\omega}{\partial\thetav}\right]^{-1}.
\end{equation}
\subsection{Design criterion}
In repeated measures trials when the interest is in only estimating direct treatment effect contrasts, we  may instead work with $Var(\hat{\tauv})$  given by,
\begin{equation}\label{pobj}
Var_\zeta(\hat{\tauv})=EVar_\zeta(\hat{\thetav})E',
\end{equation}
where $E$ is a $t\times m$ matrix given by $[0_{t1},0_{tp},I_t,0_{tt}]$ and $m$ is the total number of parameters in $\thetav$. Here by $0_{p_1p_2}$ we mean a $p_1\times p_2$ matrix of zeros.

The design minimizing the criterion
\begin{equation}
\Lambda(\zeta,\thetav, \alpha) =\log\text{Det}(Var_\zeta(\hat{\tauv})).
\end{equation}
is known as the $D_A$-optimal design (\cite{Atkinson2007} , page 137). Since it is a GLM the variance depends on the model parameters as well as the covariance parameters, and the design obtained is locally optimal.

To obtain $D_A$-optimal designs robust to uncertainties in the parameters we propose a Bayesian approach. This method has been used before for  logistic regression by \cite{Chaloner1989}, and \cite{Dror2006} and for block designs by \cite{Woods2011}. For repeated measures models, the design which minimizes
\begin{equation}\label{objectivefn}
 \Psi(\mathfrak{B},\zeta,\alpha) = \int_{\mathfrak{B}}\Lambda(\zeta,\thetav, \alpha)\,dF(\thetav),
\end{equation}
where $\mathfrak{B}\subset\mathbb{R}^{m}$ is the parameter space of  parameter vector $\thetav$ and $F(\thetav)$ is a proper prior distribution for $\thetav$, is the $D_A$-optimal Bayesian crossover design (or the average $D_A$-optimal design of \cite{Pettersson2005}). Note, for all working examples (in Sections 5.1, 5.2 and 5.3) no  prior distributions are assigned to the correlation parameters $\alpha$, designs are obtained only for some fixed values chosen for $\alpha$. However, in Section \ref{alphaprior} we investigated the design performance when there are priors on $\alpha$.

In our computations we have used both uniform and normal priors for $\thetav$. The minimization of the objective function in  (\ref{objectivefn}) with respect to $\zeta$, requires high-dimensional integral calculation. Similar to  \cite{Woods2011},  Latin Hypercube Sampling (LHS) has been used for deriving an approximate solution of the above optimization problem.

For evaluating the performance of design $\zeta$ with respect to the reference design $\zeta^*$ ($D_A$-optimal Bayesian design), we use an efficiency criterion defined as:
\begin{equation}\label{eff}
Eff_D(\zeta,\zeta^{*},\mathfrak{B},\alpha) = \left[\mbox{exp}\left\{\Psi(\mathfrak{B},\zeta^{*},\alpha)-\Psi(\mathfrak{B},\zeta,\alpha)\right\}\right]^{1/m},
\end{equation}
here $m$ is the number of model parameters. Similar efficiency function has been used before by  \cite{woods2006}.

Working correlation matrix structures such as the compound symmetric (or equi-correlated) and the AR(1) are investigated.
Under the equi-correlated covariance structure,
$R_j=(1-\alpha)I_p+\alpha J_p$,
and under the AR(1) assumption,
$
R_j=\alpha^{|i-i'|},\,i\neq i'.
$

\section{Examples}
\subsection{Example 1: Four periods, four treatments binary  response trial}
In Case study 1,  a four periods four treatments crossover trial described in \cite{Kenward1992} is considered. There are eighty subjects allocated to the four treatment sequences, with about twenty subjects per sequence. Treatments are denoted by $A$, $B$, $C$ and $D$. The treatment sequences form a \textit{Williams design} given as follows:

\begin{equation*}
 \begin{bmatrix}
A& B &C& D \\
 B&D&A&C\\
C&A&D&B\\
D&C&B&A
 \end{bmatrix}
\end{equation*}
The response variable is binary in nature. The data set is available in Table 3 of \cite{Kenward1992}.
For a four periods, four treatments trial, there are 24 possible \textit{Latin square designs} (LSDs)   with  every treatment  represented once and only once in each row and in each column (see Table 5.1 \cite{Senn_2002}). A special form of \textit{Latin square design} is called \textit{Williams square design} (WSD) in which every treatment follows every other treatment only once. In the case of normal responses when $t = p$ and $t$ is even, for reduced models (no carryover effects) LSD and  for full models (carryover effects present) WSDs  are  variance balanced designs (\cite{John2014}, page 361). However, these designs may not be optimal in general. But under some subject constraints WSD is universally optimal for even $t$, $n\leq t(t+2)/2$ and $4\leq t\leq12$ (\cite{Bose2009}, page 40).

Instead of using equation (2) directly as the linear predictor $\eta_{ij}$ we use a reparametrized version ,
\begin{equation}
\eta_{ij} = \nu + \beta_{1}^{*}P_{1} + \beta_{2}^{*}P_{2} + \beta_{3}^{*}P_{3} + \tau_{1}^{*}T_{1} + \tau_{2}^{*}T_{2}+\tau_{3}^{*}T_{3} + \gamma_{1}^{*}C_{1}+ \gamma_{2}^{*}C_{2}+ \gamma_{3}^{*}C_{3},
\end{equation}
where
\begin{table}[H]
\begin{center}
\begin{tabular}{c|cccccccc}
&$P_{1}$&$P_{2}$&$P_{3}$\\
\hline
Period 1 &0&0&0\\
Period 2 &1&0&0\\
Period 3 &0&1&0\\
Period 4 &0&0&1\\
\hline
\end{tabular}\label{partable}
\end{center}
\end{table}
$T_{i}$'s and $C_{i}$'s for $i = 1,\cdots, 4$, are similarly defined. Also, $\beta_1=0$, $\beta_i=\beta^{*}_{i-1},\,i=2,\ldots,4$, $\tau_{A}=0$, $\tau_{B}=\tau^{*}_{1}$, $\tau_{C}=\tau^{*}_{2}$, $\tau_{D}=\tau^{*}_{3}$ , $\gamma_A=0$, $\gamma_B=\gamma^{*}_{1}$, $\gamma_C=\gamma^{*}_{2}$ and $\gamma_D=\gamma^{*}_{3}$.  Note that  carryover effect in the first period is taken to be zero.  It is noted that total number of parameters reduces to $m^{*} = m-3$, where $m$ in equation (2) was 13 for a $4\times 4$ design. The $E$ matrix defined in equation (\ref{pobj})  will be of same form but $m$ is replaced by $m^{*}$.

Point estimates and corresponding confidence intervals of the parameters are calculated using \textit{PROC GENMOD} procedure in SAS software (\cite{SAS1999}). Results are summarized in Table \ref{estimatebin} for both reduced and full models. In a reduced model it is assumed that there are no carryover treatment effects, while in a full model both direct and carryover treatment effects are assumed to be present. The working correlation structure is taken to be compound symmetric (CS) in nature,  the correlation coefficient is estimated to be 0.215.

\begin{table}[H]
\caption{Point estimates and confidence intervals for both reduced and full models under the compound symmetric correlation structure (Example 1).}
\begin{center}
\begin{tabular}{c|cccccccccc}
Parameter & \multicolumn{2}{c}{ Point estimate [95\%Confidence interval]}\\
\hline
\mbox{}& no carryover effect & with carryover effect\\
\hline
$\nu$ & 1.0980 [0.4232\;1.7728] & 1.0158  [0.3474\; 1.6842]\\
$\beta_{1}^{*}$ & -0.3056 [-0.8643\;0.2532]&  -0.5525  [-1.2565\; 0.1515]\\
$\beta_{2}^{*}$ & -0.2414 [-0.8228 \;0.3399]& -0.4842  [-1.2034\; 0.2349]\\
$\beta_{3}^{*}$ &  0.3817 [-0.2391 \;1.0026]& 0.1234  [-0.6888\; 0.9356]\\
$\tau_{1}^{*}$  &-0.3270 [-0.8660\;0.2119]& -0.2564  [-0.8075\; 0.2948]\\
$\tau_{2}^{*}$ &  -0.0681 [-0.6996\;0.5635]&  0.0069  [-0.6473\; 0.6610]\\
$\tau_{3}^{*}$ &  -0.5322 [-1.1684\;0.1041]&  -0.3736  [-1.0165\; 0.2693]\\
$\gamma_{1}^{*}$ &-& 0.1786  [-0.5965\; 0.9538]\\
 $\gamma_{2}^{*}$ &-& 0.2242  [-0.5443\; 0.9927]\\
 $\gamma_{3}^{*}$ &-& 0.6620  [-0.1352\;1.4591]\\
\hline
\end{tabular}\label{estimatebin}
\end{center}
\end{table}

For a $4\times 4$ crossover trial the number of all possible treatment sequences are $4^4=256$. However, in this example we restrict our design space to only 16 treatment sequences, i.e., $\Omega=$ \{$ACDB$, $BDCA$, $CBAD$, $DABC$, $ADCB$, $BCDA$, $CABD$, $DBAC$, $AABB$, $BBAA$, $CCDD$, $DDCC$, $AAAB$, $BBBA$, $CCCD$, $DDDC$\}. These sequences are chosen since  they can be used to form LSDs (including WSDs) and also non LSD crossover designs. Note in the normal response case it has been reported that WSDs under certain constraints are universally optimal for the 4 treatment and 4 period case. Thus, we felt it was enough to restrict $\Omega$ to these 16 sequences. Also lowering the number of treatment sequences increases our computational speed. The Bayesian designs found, also satisfy the conditions of the equivalence theorem given in the appendix.

The following prior distributions are considered for the model parameters, $\thetav$, for obtaining the Bayesian optimal design:
\begin{itemize}
\item[Prior 1:] Cartesian product of 95\% confidence intervals of parameters given in Table \ref{estimatebin}.

\item[Prior 2:] Cartesian product of the nonnegative part of 95\% confidence intervals of parameters given in Table \ref{estimatebin}.

 \item[Prior 3] and 4:  Independent multivariate normal distribution with mean vector as the point estimates of the parameters given in  Table \ref{estimatebin} and (for prior 3) the variance is 0.25, (for prior 4) the variance is 0.50.
\end{itemize}
Note prior 2 is asymmetric around 0 and priors 3 and 4 are the normal priors with different variances.
The Bayesian crossover design  is obtained by minimizing formula \ref{objectivefn}, and denoting it by $D^{B}$.


The performance of $D^{B}$ is compared with  24 LSDs including 6 WSDs, and 24 extra period designs (EPDs) (a design in which first three rows correspond to a LSD and the last row is same as the previous one (\cite{Patterson1959}). We noted that the performance of each LSD is same among the 18 LSDs  under the reduced and full models for both of the correlation structures and priors used. Same is true for 6 WSDs and 24 EPDs. Thus the results are based on one LSD, one WSD and one EPD.

\subsubsection{Reduced model: No carryover effects}
The Bayesian $D_A$-optimal design is obtained under three correlation structures, independent ($\alpha=0$), compound symmetric (CS) and AR(1). The proportions assigned to each treatment sequence by $D^B$ for varying $\alpha$ are plotted in Figure \ref{exm1pw}. It is noted (see Figure \ref{exm1pw}(A)) that under the independent correlation structure (i.e., $\alpha$ = 0), $D^{B}$ utilizes all the 16 sequences for priors 1, 3 and 4. In the case of prior 2 and $\alpha=0$, the sequences $\{BDCA, CBAD, BCDA, DBAC, AAAB, BBBA\}$ are left unused. As $\alpha$ increases for the CS structure, $D^B$ utilizes the sequences forming a LSD ($ADCB, BCDA,DABC,CBAD$) with almost 100\% weightage and equal proportions to each. Under the AR(1) structure, $D^B$ uses only the first eight sequences (see Figure \ref{exm1pw}(B)). The efficiencies of the LSD, WSD and EPD with respect to $D^B$ are presented in  Figure \ref{exm1eff} (A). Note that under CS structure both LSD and WSD  designs are as good as $D^{B}$, while EPD has lower efficiency, especially for priors 1 and 2. Efficiencies of LSD and WSD are constant with respect to $\alpha$ and also overlap. Under the AR(1) structure (see Figure \ref{exm1eff} (B)),  WSD is more efficient followed by LSD, and EPD performs worst. Note that performance of  EPD  also worsens as $\alpha$ increases. Efficiency comparisons are not much affected by the choice of the priors in the AR(1) case.

\subsubsection{Full model: With carryover effect}
\begin{center}
\begin{figure}[H]
\hspace*{-0.7cm}\includegraphics[scale=0.49]{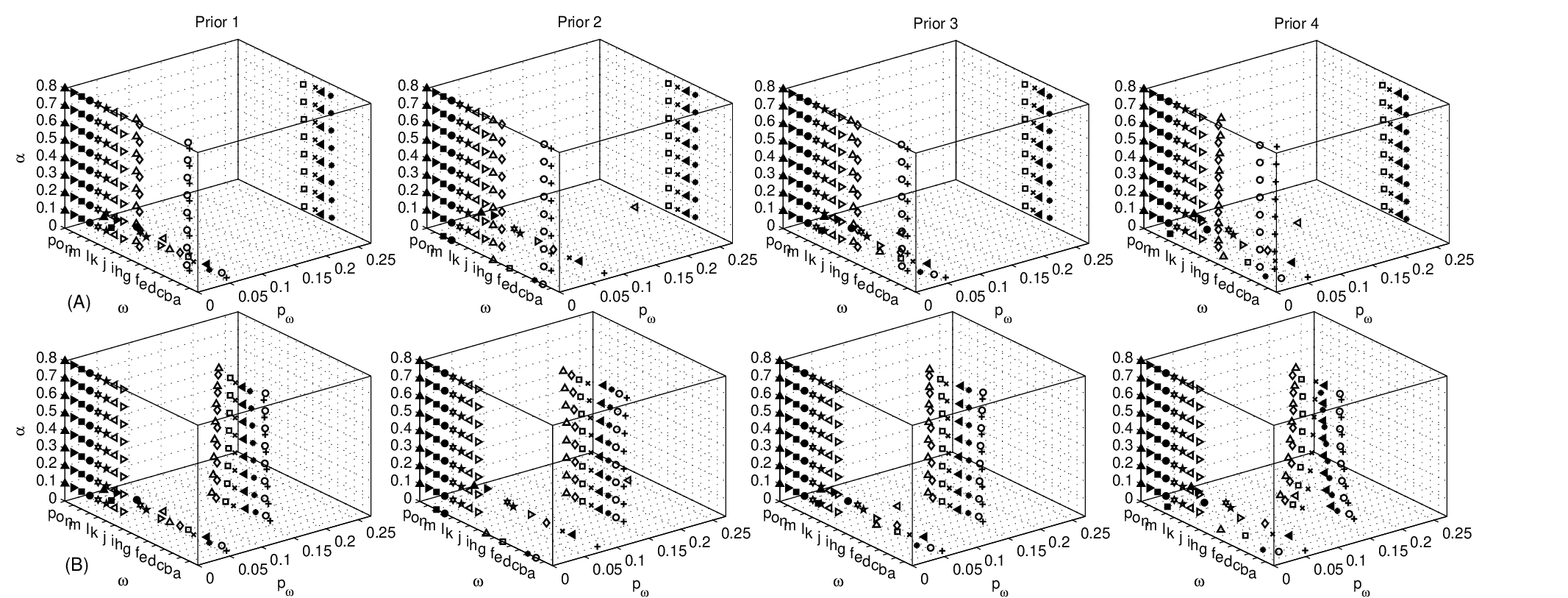}
\hspace*{-0.7cm}\includegraphics[scale=0.5]{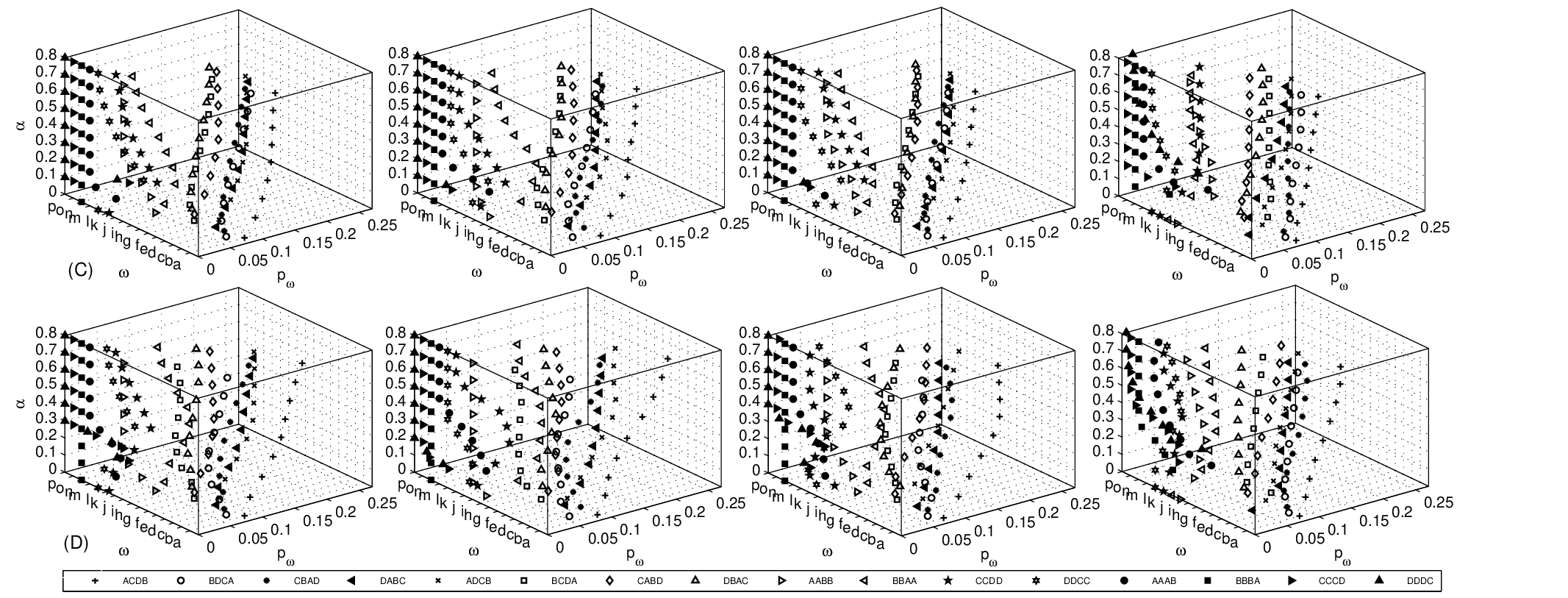}
\caption{Weights ($p_\omega$) versus the treatment sequences for different $\alpha$ values using priors 1-4 for the model parameters in Example 1. Treatment sequences  labeled as $\{a, b,...,p\}$   correspond  to the treatment sequences given in the design space $\Omega$ in Example 1 (Section 5.1). (A): Model with no carry over effect and correlation structure is CS  (B): Model with no carry over effect and correlation structure is AR(1) (C): Model with carry over effect and correlation structure is CS (D): Model with carry over effect and correlation structure is AR(1)}
\label{exm1pw}
\end{figure}
\end{center}

\begin{center}
\begin{figure}[H]
\hspace*{-0.7cm}\includegraphics[scale=0.5]{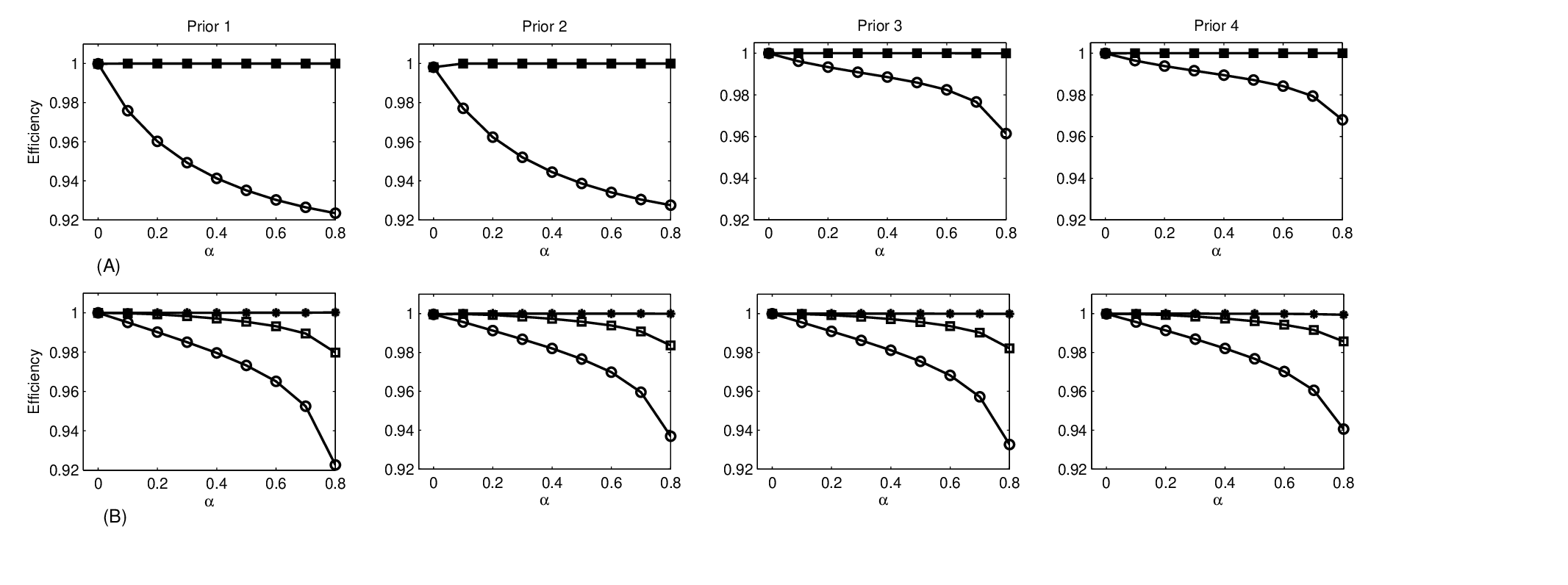}
\hspace*{-0.7cm}\includegraphics[scale=0.51]{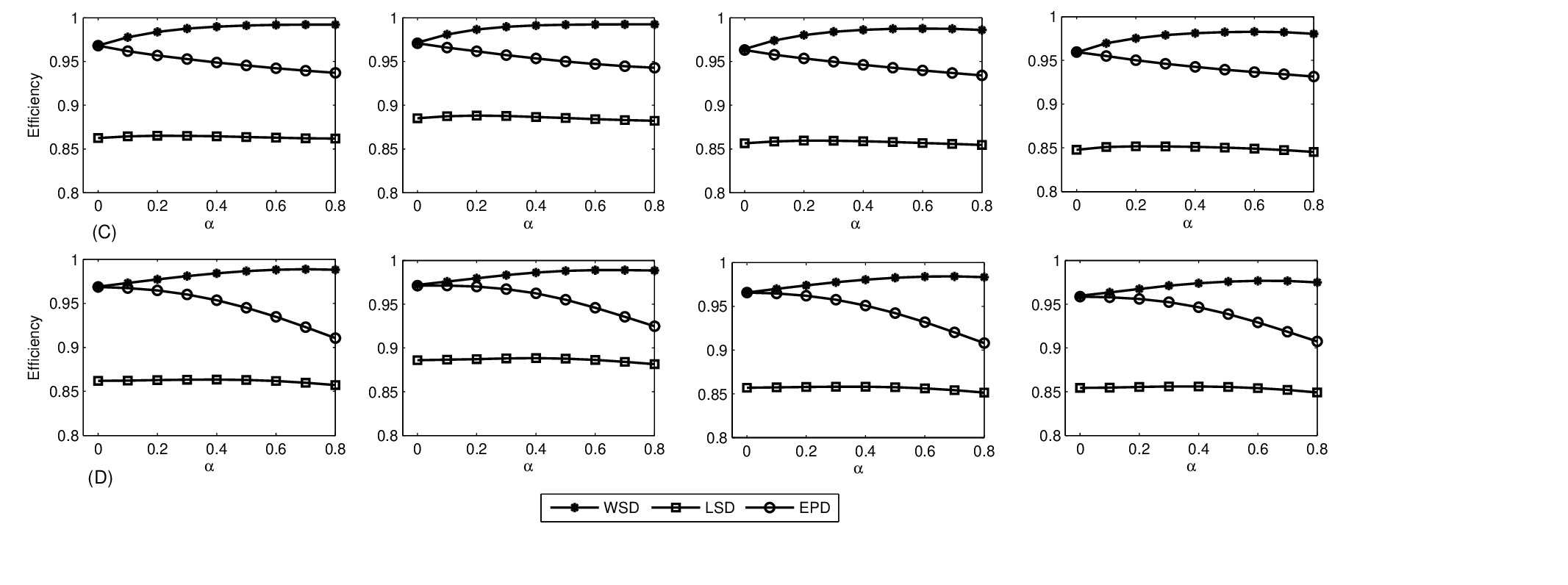}
\caption{Efficiencies of  WSD, LSD and EPD compared with $D^B$ using  priors 1-4 for the model parameters in Example 1, (A): Model with no carry over effect and correlation structure is CS  (B): Model with no carry over effect and correlation structure is AR(1) (C): Model with carry over effect and correlation structure is CS (D): Model with carry over effect and correlation structure is AR(1)}
\label{exm1eff}
\end{figure}
\end{center}

It is observed from Figures \ref{exm1pw} (C) and (D), for $\alpha = 0$, under priors 2 and 3, $D^{B}$ utilizes all sequences except $\{BBBA\}$. For $\alpha=0$, prior 1: sequences $\{BBBA, CCDD, DDCC\}$ and prior 4: $\{BBBA, CCDD, DDCC,AABB\}$, are left unused , respectively.  Under the CS and AR(1) structure, $D^B$ utilizes the first eight sequences with more than 70\% of weight (see Figure \ref{exm1pw} (C) and (D)), and as $\alpha$ increases the first eight sequences get more than 80\% weight.  It can be observed from Figure \ref{exm1eff} (C) and (D) that WSD is most efficient as compared to LSD and EPD under all correlation structures. Also contrary to the reduced model, here the  LSD performs worse (with about $85\%$ efficiency) than EPD. Efficiency comparisons are not much affected by the choice of the priors.
Equation (\ref{eqv2}) in Theorem 1 in the appendix has been used to confirm the $D_A$-optimality of all Bayesian designs obtained for both reduced and full models.

\subsection{Example 2: Two periods two treatments Poisson  response trial}
A crossover trial with two drugs given in two periods for controlling angina in 20 patients is considered. The count of attacks suffered by the patients is assumed to be a Poisson random variable.  Treatment sequences  $AB$ and $BA$ are used in the trial. However, we should note that this design does not permit the unbiased estimation of the treatment contrast under carryover effect (\cite{Jones_2014}),  though the estimates and corresponding confidence intervals may still be used to choose the prior distributions.

Reparametrizing the linear predictor $\eta_{ij}$ for this $2\times 2$ crossover design  as done by Laska and Meisner (1985),
\begin{equation*}
\eta_{ij} = \nu + \beta^{*} P +  \tau^{*} T+ \gamma^{*} C.
\end{equation*}
Here, $\tau^*=(\tau_A-\tau_B)/2$, $\gamma^*=(\gamma_A-\gamma_B)/2$, $\beta_1=0$  and $\beta_2=\beta^{*}$. The variables $P$ is coded 1 for period $2$ and zero otherwise, while $T,C=1$ for treatment $A$ and $-1$ for treatment $B$. It is assumed that carryover effect is zero in the first period. For a $2\times2$ cross-over trial  compound symmetric and AR(1)  correlation structures are equal. Estimation of the parameters is again done by using \textit{PROC GENMOD} in SAS software (\cite{SAS1999}). Point estimates and their 95\% confidence intervals are listed in Table \ref{estimateexm1}. Estimate of the correlation coefficient is $\alpha = 0.0798$.
\begin{table}[H]
\caption{Point estimates and confidence intervals for both reduced and full models for Poisson data in Example 2.}
\begin{center}
\begin{tabular}{c|cccccccccc}
Parameter & \multicolumn{2}{c}{ Point estimate [95\% Confidence interval]}\\
\hline
\mbox{}& no carryover effect & with carryover effect\\
\hline
$\nu$ & 0.0493 [-0.4457 \;0.5444]&-0.0541 [-1.0405\; 0.9324] \\
$\beta^{*}$ & -0.0011 [-0.4256 \;0.4234]&0.0541 [-0.4519\; 0.5600]\\
$\tau^{*}$ &  0.5664 [0.1006\;1.0322]&0.6419 [-0.1036\; 1.3873]\\
$\gamma^{*}$ &-&0.1494 [-0.8566\; 1.1553]\\
\hline
\end{tabular}\label{estimateexm1}
\end{center}
\end{table}

For a $2\times 2$ crossover design, the set of all possible treatment sequences is taken to be $\Omega=$ \{$AB,BA,AA,BB$\}. The Bayesian design with the $D_A$-optimal allocation of subjects to the treatment sequences \{$AB,BA,AA,BB$\} is denoted by $D^B$.  The performance of $D^B$  is compared to  $D_{I}=\{AB,BA,AA,BB\}$ and  $D_{II}=$ \{$AB,BA$\}. Both $D_I$ and $D_{II}$ assigns equal allocation to each of their treatment sequences.

Following prior distributions for the model parameters were chosen:
\begin{itemize}
\item[Prior 1:] Cartesian product of 95\% confidence intervals of parameters given in Table \ref{estimateexm1}.

\item[Prior 2:] Cartesian product of the nonnegative part of 95\% confidence intervals of parameters given in Table \ref{estimateexm1}.

 \item[Prior 3] and 4:  Independent multivariate normal distribution with mean vector as the point estimates of the parameters given in  Table \ref{estimateexm1} and (for prior 3) the variance is 0.25, (for prior 4) the variance is 0.50.
\end{itemize}

\subsubsection{Reduced model: No carryover effects}
By observing Figure \ref{exm2pw} (A), it is concluded that for $\alpha = 0$, $D^{B}$ utilizes all  sequences, for all priors. For $\alpha>0$, the Bayesian crossover design $D^B$ for the reduced model consists of sequences \{$AB$, $BA$\} with approximately equal weightage to each sequence, thus $D_{II}$ and $D^B$ are very similar under the reduced model. From Figure \ref{exm2pw} (C) we see that $D_{II}$ is more efficient than $D_{I}$ and also the performance of $D_{I}$ worsens as  $\alpha$ increases. Choices of the prior distributions do not effect the results. Also, the results matches with those for the normal response model for a $2\times 2$ crossover design (\cite{Laska1985}).

\subsubsection{Full model: With carryover effects}
\begin{center}
\begin{figure}[H]
\includegraphics[scale=0.5]{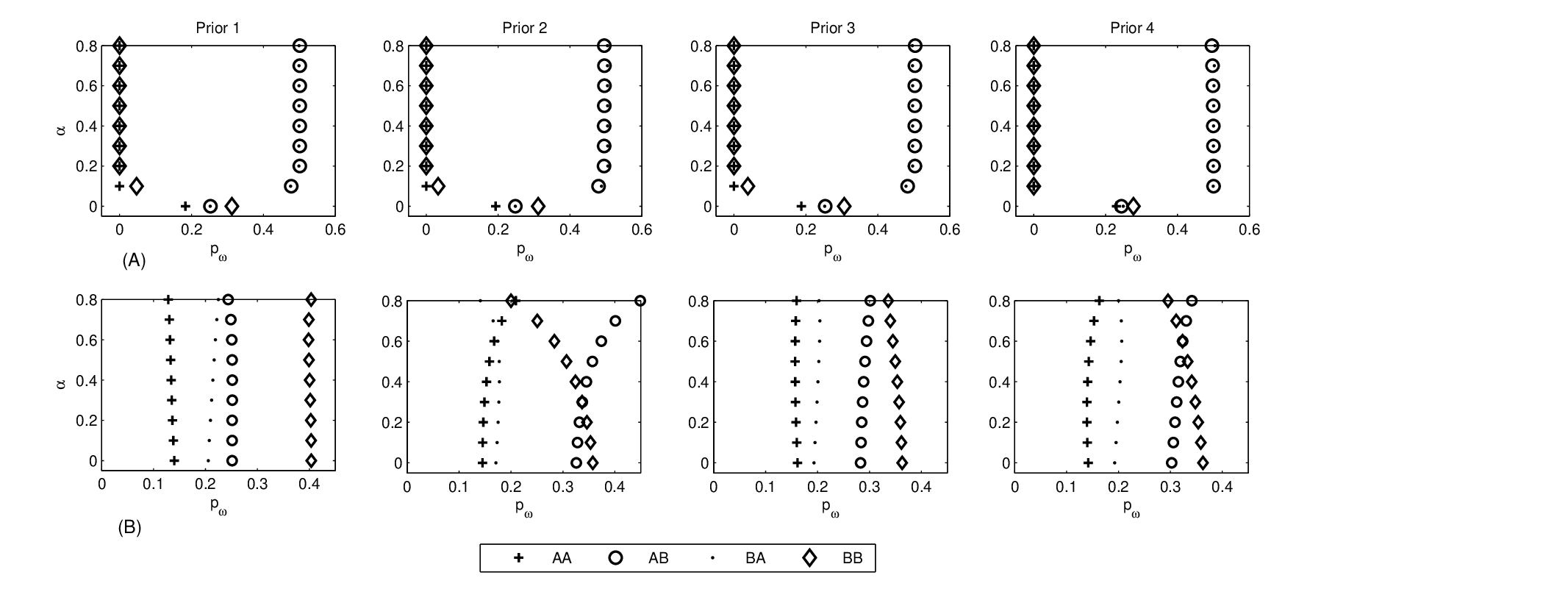}
\includegraphics[scale=0.5]{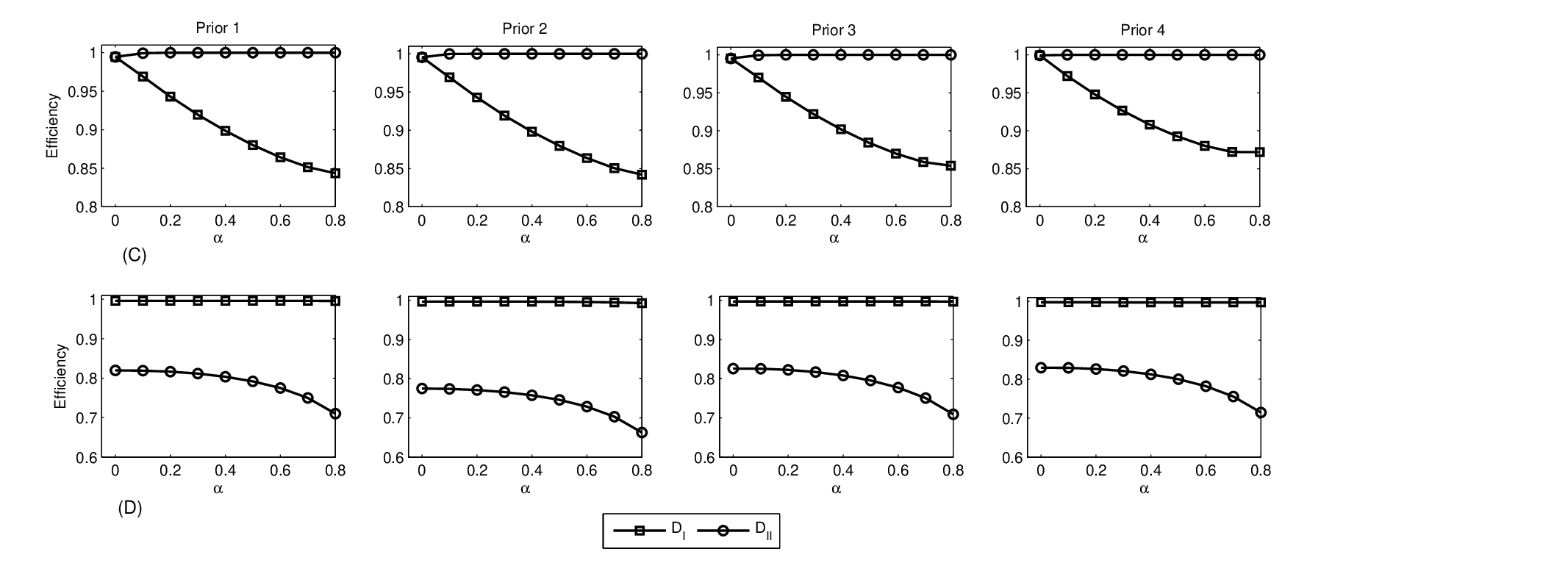}
\caption{Weights ($p_\omega$) assigned to each of  the treatment sequences for different $\alpha$ values  using priors 1-4 for the model parameters in Example 1 in  (A): Model with no carry over effect  (B): Model with carry over effect.   Efficiency plots of designs $D_{I}$ and $D_{II}$ in (C): Model with no carry over effect  (D): Model with carry over effect}
\label{exm2pw}
\end{figure}
\end{center}

Introducing crossover effects in the model, however changes the results completely except for the $\alpha=0$ case. The Bayesian crossover design $D^B$ for the full model now utilizes the sequences \{$AA$, $AB$\} and its dual.  Proportions assigned to the treatment sequences are sensitive to the choice of  the prior distribution as noted from Figure \ref{exm2pw} (B). Figure \ref{exm2pw} (D) shows that the design $D_{I}$ has efficiency values close to 1 and performs better than $D_{II}$. Also,  $D_{II}$  is  affected by increasing $\alpha$ (see Figure \ref{exm2pw} (D)). For normal responses in case of a full model, similar results are noted by (\cite{Laska1985}).
All designs obtained for the Poisson response here are verified to be $D_{A}$-optimal  using Theorem \ref{thm1} given in the Appendix.

\subsection{Example 3: Three periods two treatment Gamma response trial}
We consider a hypothetical gamma response trial with two treatments, $A$ and $B$ applied in three periods. For the $3\times 2$ crossover design the set of all possible treatment sequences is taken to be  $\Omega=\{AAA, AAB, ABB, ABA, BBA, BAA, BAB, BBB\}$. The response is length of hospital stay which is assumed to follow a Gamma distribution. The Bayesian crossover design $D^{B}$ is determined by searching over $\Omega$. The linear predictor is again reparametrized as in Example 2, using $\tau^*=(\tau_A-\tau_B)/2$, $\gamma^*=(\gamma_A-\gamma_B)/2$, $\beta_1=0$, $\beta_{i}=\beta^{*}_{i-1},\,i=2,3$, and $T,C=1$ for treatment $A$ and $-1$ for treatment $B$,
\begin{equation*}
  \eta_{ij} = \nu + \beta_{1}^{*}P_{1} + \beta_{2}^{*}P_{2} + \tau^{*} T + \gamma^{*} C,
\end{equation*}
it is assumed that carryover effect is zero in the first period.

The data sets are simulated using the parameter values $(\nu, \beta_{1}^{*},\beta_{2}^{*}, \tau^{*}) = (0.50, 0.15, 0.20, 0.25)$ for a reduced model and $(\nu, \beta_{1}^{*},\beta_{2}^{*}, \tau^{*}, \gamma^{*}) = (0.50, 0.20, 0.30, 0.25, 0.15)$ for a full model. We have considered the treatment sequences $ABB$ and $BAA$ with the assignment of 10 subjects each to generate the data. Observations are assumed to be independent within the periods. The link function used is the reciprocal link and the shape parameter $\kappa$ is fixed at 2.0. Point estimates and corresponding confidence intervals of the parameters are calculated using \textit{PROC GENMOD} procedure in SAS software (\cite{SAS1999}).

Following prior distributions for the model parameters are used to obtain the Bayesian optimal designs:
\begin{itemize}
\item[Prior 1:] Cartesian product of 95\% confidence intervals of parameters given in Table \ref{estimateP}.

\item[Prior 2:] Cartesian product of the nonnegative part of 95\% confidence intervals of parameters given in Table \ref{estimateP}.

 \item[Prior 3:] $(\nu, \beta_{1}^{*},\beta_{2}^{*}, \tau^{*}, \gamma^{*})\in [-100,\;100]\times [-100,\;100]\times [-100,\;100]\times [-100,\;100]\times [-100,\;100]$.
 \item[Prior 4] and 5:  Independent multivariate normal distribution with mean vector as the point estimates of the parameters given in  Table \ref{estimateP} and (for prior 4) the variance is 0.25, (for prior 5) the variance is 0.50.
\end{itemize}
For the reciprocal link function, we use the restriction  $\eta_{ij}>0, i = 1,\cdots,p, j = 1,\cdots,n$. Prior 3 is a new prior considered here. Priors similar to prior 3 were not used in Examples 1 and 2, since such large values of parameters may have introduced singularity in the asymptotic variance covariance matrix of the parameter estimates.

The Bayesian design $D^{B}$ has been compared with the following designs:
\begin{itemize}
\item[$D_{a}$] = $ABB, BAA, AAB, BBA$ with equal allocation to each treatment sequence.
\item[$D_{b}$] = $ABB, BAA$ with equal allocation to each treatment sequence.
\item[$D_{c}$] = $ABA, BAB, ABB, BAA$ with equal allocation to each treatment sequence.
\end{itemize}

\begin{table}[H]
\caption{Point estimates and confidence intervals for both reduced and full models for Gamma response in Example 3.}
\begin{center}
\begin{tabular}{c|cccccccccc}
Parameter & \multicolumn{2}{c}{Point estimate [95\% Confidence interval]}\\
\hline
\mbox{}& no carryover effect & with carryover effect\\
\hline
$\nu$ &   0.5846 [ 0.3137\;   0.8556]&   0.4653 [ 0.2671\;   0.6635] \\
$\beta_{1}^{*}$ &  0.1842 [ -0.0906\;   0.4591] &   0.1360 [ -0.1814\;   0.4535]\\
$\beta_{2}^{*}$ &  0.2422 [ -0.0873\;   0.5717] &   0.3661 [ 0.0818\;   0.6503]\\
$\tau^{*}$ &    0.2310 [ 0.0446 \;  0.4173] &   0.2830 [ -0.0150\;   0.5810]\\
$\gamma^{*}$ &-&   0.1178 [ -0.3020\;   0.5377]\\
\hline
\end{tabular}\label{estimateP}
\end{center}
\end{table}

\subsubsection{Reduced model: No carryover effects}
 For log link function, asymptotic variance covariance matrix of parameter estimates does not depend on the model parameters, $\thetav$, as observed from the information matrix given in section \ref{gamma} under the log link function. Thus results are similar to those in the normal response case. Under the CS structure, design \{ABB, BAA\} with equal proportions is the optimal design for direct treatment effect.  Under AR(1) structure, optimal design utilizes the sequences $ABA$ and $BAB$ with equal proportions.

For reciprocal link function (see Figure \ref{exm3pw} (A)), it is seen that for $\alpha=0$, all 8 sequences are utilized by $D^B$ except in the case of prior 3 (sequences $BAB, ABB$ are not used). For most of the positive $\alpha$ values and priors 1, 2 and 5, for the CS structure  $D^B$ utilizes the sequences $\{BBA, BAB, BAA, ABB\}$. For priors 3 and 4 the sequence $BAA$ is left unused for high $\alpha$ values. Under each prior approximately 40\% weightage is given to sequence $ABB$. With an increase in $\alpha$, weights on $BBA$ and $BAB$ also increase however weights on $BAA$ decrease. We also note that the weights on $BAA$ is sensitive to the prior used. From Figure \ref{exm3eff} (A), it is noted that $D_b$ is most efficient, this is also true for normal responses.

Under AR(1) structure, $D^B$ utilizes the sequences $\{ABA, BAB\}$ with approximately 45\% and 55\% weightage, respectively, for each priors.  These proportions are not affected by increasing  $\alpha$ values (see Figure \ref{exm3pw} (B)). From the efficiency plots (see Figure \ref{exm3eff} (B)) design $D_c$ turns out to be the most efficient as compared to $D_a$ and $D_b$.
\subsubsection{Full model: With carryover effects}

For log link function again the results are similar to those in the normal response case. Under CS structure, $\{ABB, BAA\}$ is the optimal design and for AR(1) structure, optimal design is $\{ABB, AAB, BAA, BBA\}$ with more than 90\% weightage given  to the sequence $AAB$ and its dual.

\begin{center}
\begin{figure}[H]
\hspace*{-0.7cm}\includegraphics[scale=0.5]{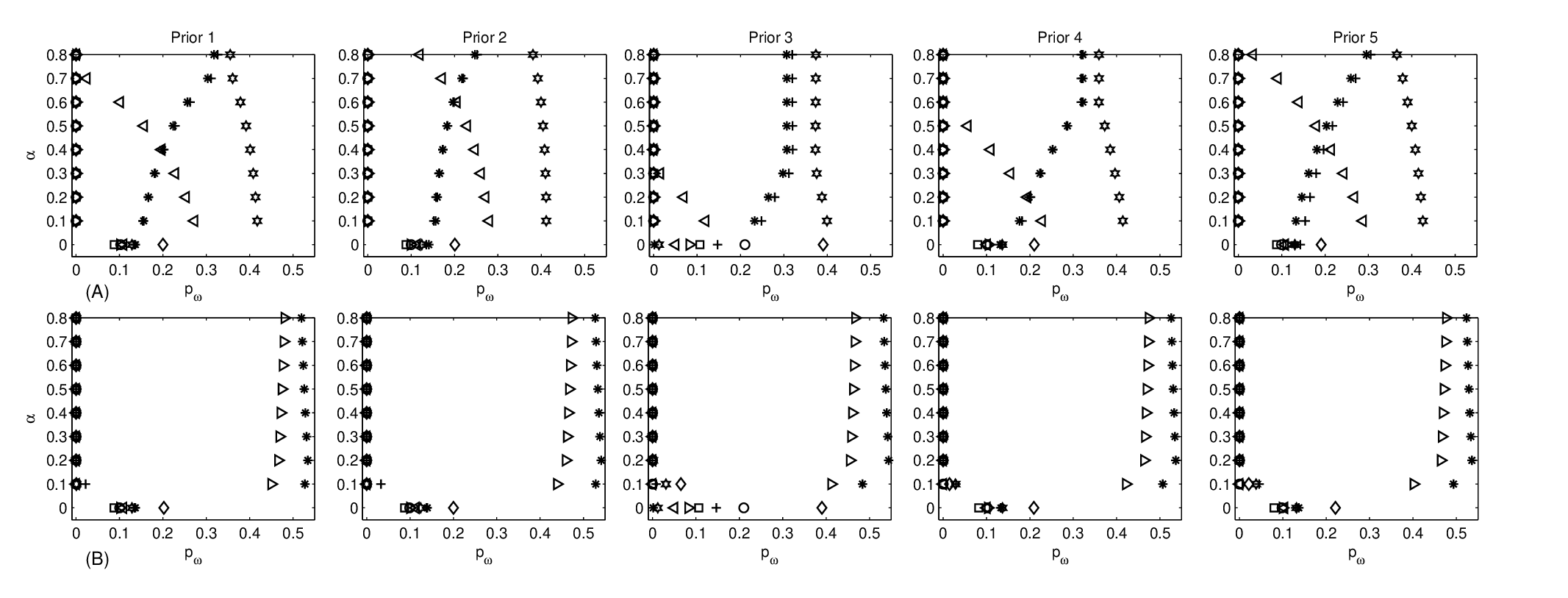}
\hspace*{-0.7cm}\includegraphics[scale=0.5]{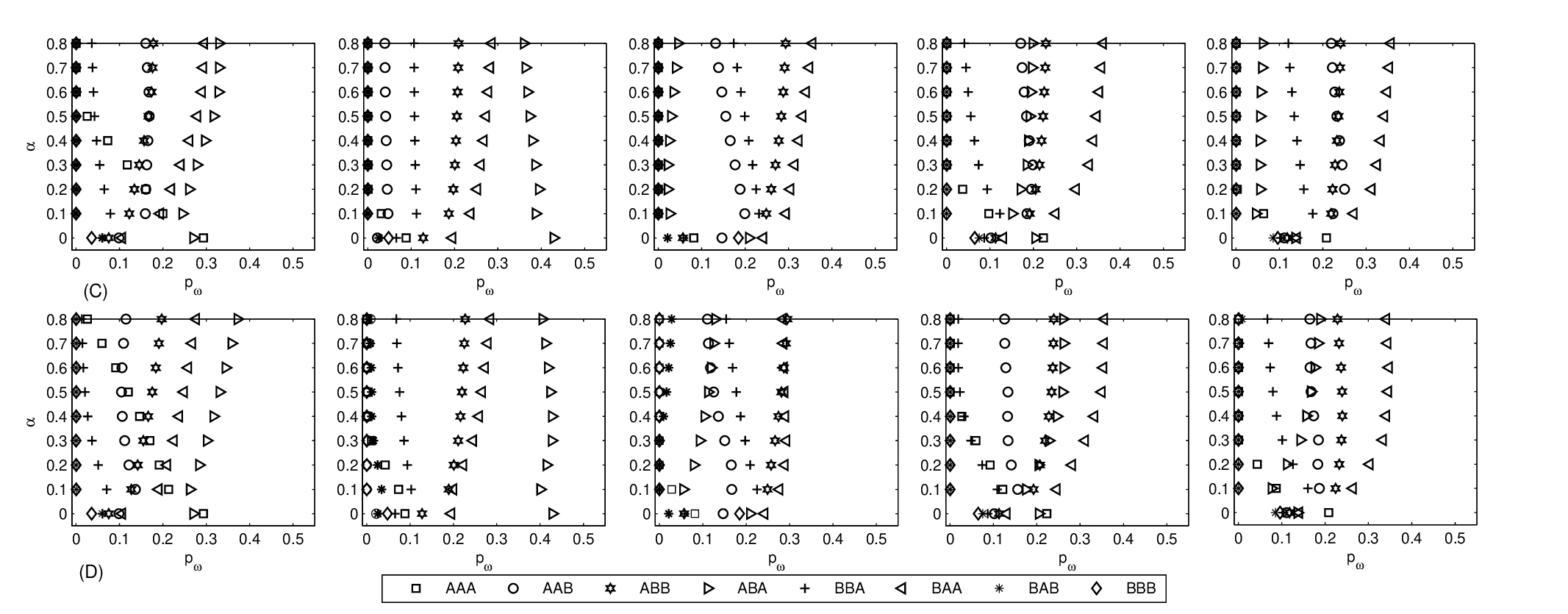}
\caption{Weights  ($p_\omega$) assigned to each treatment sequence for  different $\alpha$ values using priors 1-5 for the model parameters in Example 3, (A): Model with no carry over effect and correlation structure is CS  (B): Model with no carry over effect and correlation structure is AR(1) (C): Model with carry over effect and correlation structure is CS (D): Model with carry over effect and correlation structure is AR(1)}
\label{exm3pw}
\end{figure}
\end{center}

\begin{center}
\begin{figure}[H]
\hspace*{-1cm}\includegraphics[scale=0.5]{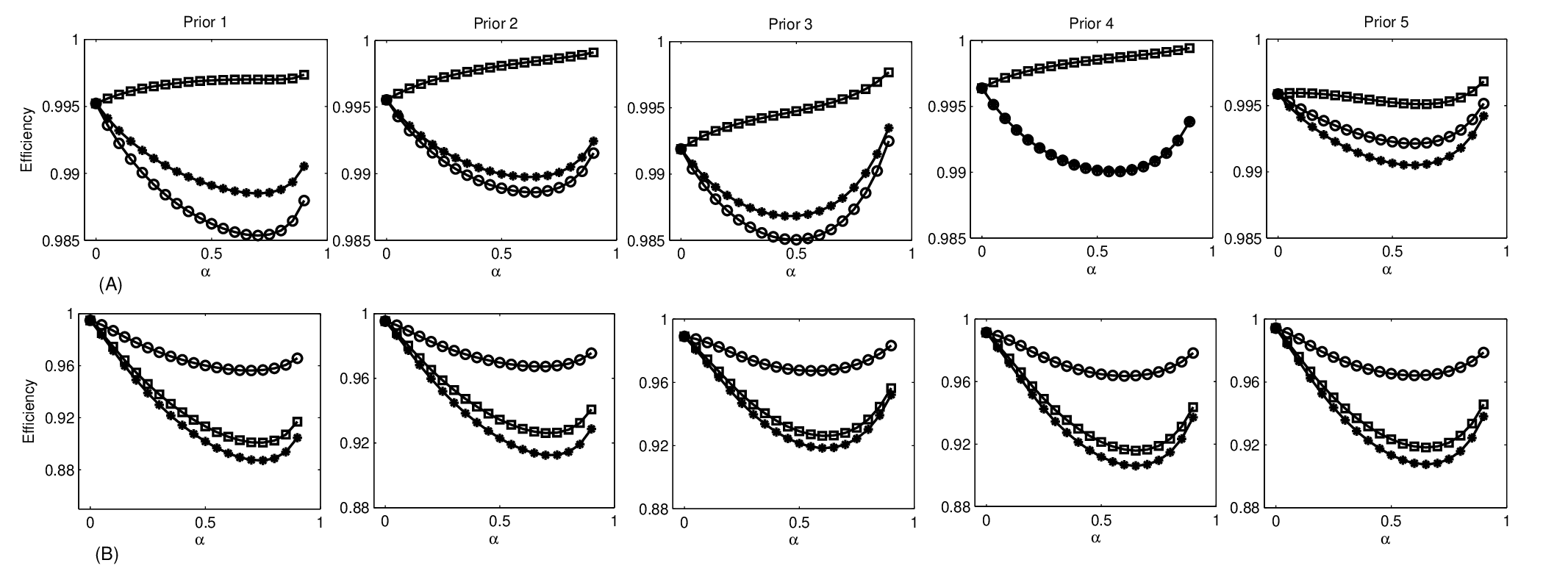}
\hspace*{-1cm}\includegraphics[scale=0.5]{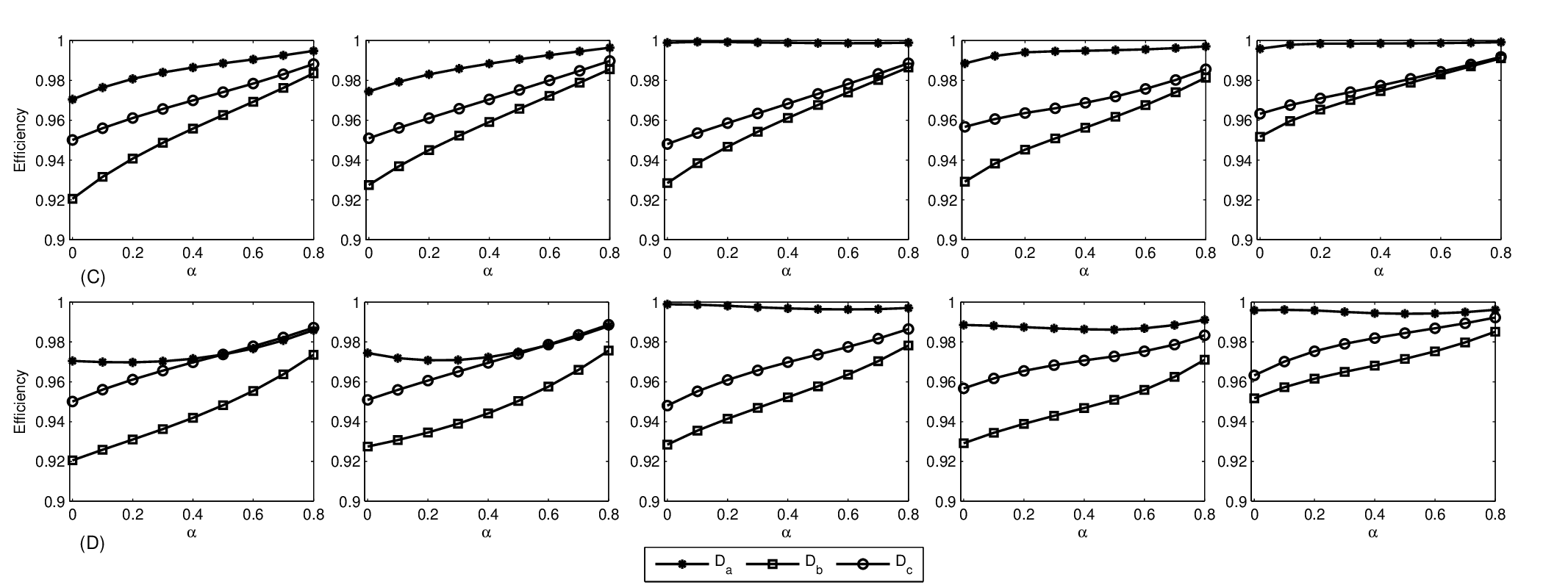}
\caption{Efficiencies of designs $D_a$, $D_b$ and $D_c$ compared with $D^B$  using priors 1-5 for the model parameters in Example 3, (A): Model with no carry over effect and correlation structure is CS  (B): Model with no carry over effect and correlation structure is AR(1) (C): Model with carry over effect and correlation structure is CS (D): Model with carry over effect and correlation structure is AR(1)}
\label{exm3eff}
\end{figure}
\end{center}

For reciprocal link function, when $\alpha=0$, $D^B$ uses all sequences.  Under both CS and AR(1) structures, $D^B$ uses the sequences $\{AAB, BAA, ABB, BAA, ABA\}$ (see Figure \ref{exm3pw} (C) and (D)). It is observed that for smaller values of $\alpha$, the treatment sequence $AAA$ is  included in the design. In Prior 1, the treatment sequence $AAA$ has  approximate 30\%  weight for $\alpha = 0$ and weightage decreases as $\alpha$ increases.
From the efficiency plots (Figure \ref{exm3eff} (C) and (D)), observe that design $D_a$ is the most efficient for CS correlation structures. Under the AR(1) structure, again $D_a$ performs well compared to other designs. Under prior 1 and 2, design $D_{c}$ has approximate equal efficiency as $D_{a}$ for $\alpha>0.4$.  Note that, design $D_a$ is the optimal design for normal responses under AR(1) as noted in \cite{Laska1985}.

Note again all  designs obtained in this section are verified to be optimal using Theorem \ref{thm1} given in the Appendix.

\section{Sensitivity of designs  to the assumed correlation structure}
Till now in all our computations we assume that the working correlation (WC) matrix is equal to the true correlation (TC) matrix as defined in equation (\ref{infmat}). In this section we  investigate the effect of misspecifying the correlation on performance  of designs. For illustration, Example 1 is used. There are two cases considered: Case (1): the working correlation structure is  compound symmetric but  the true correlation structure is  AR(1), Case (2): the working correlation structure is  AR(1) but  the true correlation structure is  compound symmetric. Prior 1 used before in Example 1 is assigned to the regression parameters.

\begin{center}
\begin{figure}[H]
\includegraphics[scale=0.5]{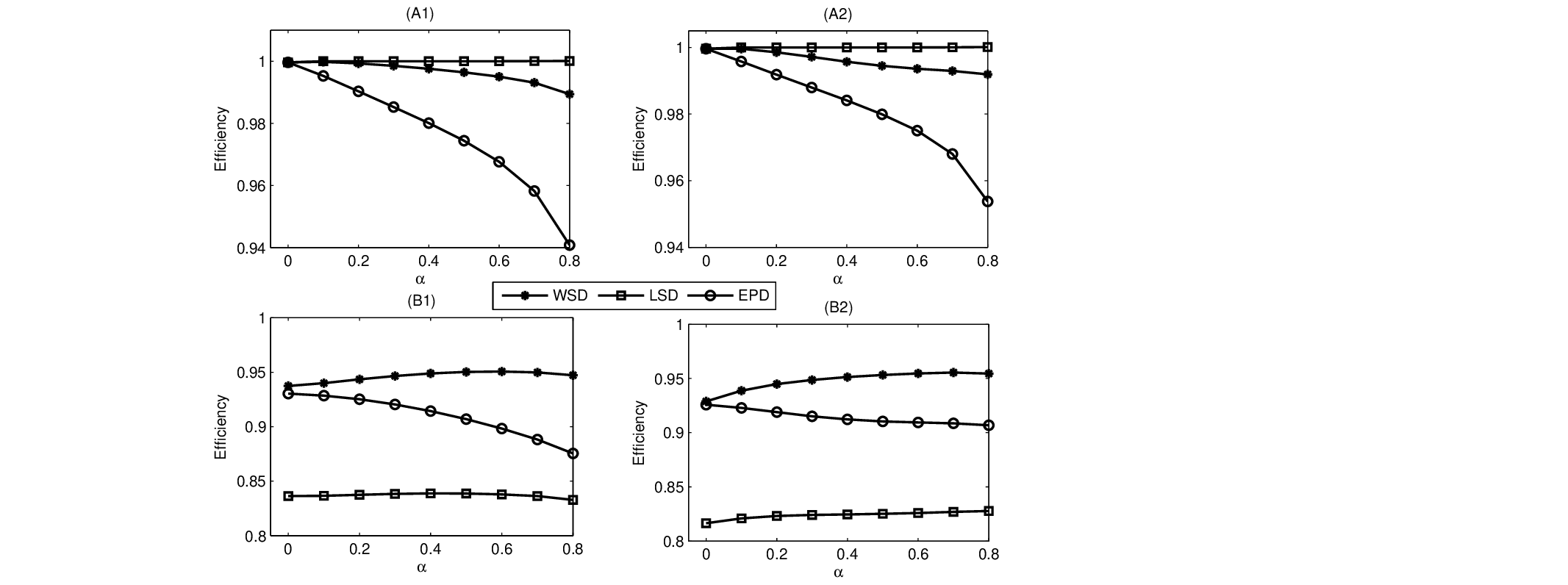}
\caption{Efficiency plots of WSD, LSD and EPD compared to $D^B$ when true correlation (TC) and working correlation (WC) are not equal (\textbf{A1}) Model without carryover effect and WC = compound symmetric (CS), TC = autoregressive (AR(1)), (\textbf{A2}) Model without carryover effect and WC = AR(1), TC = CS, (\textbf{B1}) Model with carryover effect and WC = CS, TC = AR(1),  (\textbf{B2}) Model without carryover effect and WC = AR(1), TC = CS}
\label{exm1missp}
\end{figure}
\end{center}

First we consider a model without the carryover effect. The Bayesian $D_A$-optimal design, $D^B$, is found using equation  (\ref{optcrt}) in equation (\ref{objectivefn}). $D^B$ utilizes the sequences forming a LSD under both cases 1 and 2.   Under misspecification, performance of WSD is affected very slightly (see Figure \ref{exm1missp} (A1) and (A2)), EPD performs the worst and its performance worsens with $\alpha$.. However, for the TC=WC case, we had noted earlier that  both LSD and WSD are equally efficient. Thus, misspecification under the reduced model case, has a slight adverse effect on the performance of the WSD  but not the LSD.

For the model with carryover effect, for both cases 1 and 2, $D^B$ utilizes the first 8 sequences with more than 70\% weights (this is consistent with results obtained under TC = WC). Though the performance of WSD is affected it is still the  most efficient compared to EPD and LSD (see Figure \ref{exm1missp} (B1) and (B2)), while LSD is the worst.

\section{Prior distributions on $\alpha$ and its effect on design performances}\label{alphaprior}
Designs obtained so far are based on some fixed values of  $\alpha$.  In this section, we validate the performance of the optimal designs using priors on $\alpha$. For illustration purpose, we consider the two periods two treatments Poisson  response model with carry over effect from Example 2. Prior 1 of Example 2 is chosen for the parameters involved in the linear predictor. The estimate of $\alpha$ using the data given in Example 2 is 0.0798. Based on this information we use the following  set of priors covering the value 0.0789.
\begin{itemize}
\item (i) Uniform(0, 0.2) (ii) Uniform(0, 0.5) (iii) Uniform(0, 0.8) (iv) Uniform(0, 1)

\item (i) Beta(2, 38) (ii) Beta(4, 12) (iii) Beta(6, 10) (iv) Beta(5, 5).
\end{itemize}

The first four uniform and beta priors (i-iv) are chosen such that they have similar ranges, i.e., the range of uniform (i) is similar to beta prior (i) and so on. They are also chosen to look at the effect of increasing uncertainty of the prior information on the designs.  Uniform and Beta priors have been used before by \cite{Spiegelhalter_2001} and \cite{Singh2016} for the correlation parameter of cluster randomized trials.

The ${D_A}$-optimal Bayesian criterion defined in equation (\ref{objectivefn}) changes to the design which minimizes

\begin{equation}\label{objectivefnal}
 \Psi(\mathfrak{B^{*}},\zeta) = \int_{\mathfrak{B^{*}}}\Lambda(\zeta,\thetav, \alpha)\,dF(\thetav,\alpha),
\end{equation}
where $\mathfrak{B^{*}}\subset\mathbb{R}^{m}\times[0, 1]$ is the parameter space of  parameter vector $(\thetav, \alpha)$ and $F(\thetav, \alpha)$ is a proper prior distribution for $(\thetav, \alpha)$. Optimal proportions of $D^B$ using the above criterion for different priors of $\alpha$ are given in Table \ref{missptable}. As noted before in Example 2, design $D_{I} = \{AA, AB, BA, BB\}$ with equal proportions performs well as compared to the Bayesian $D_A$-optimal design with efficiency values approximately equal to 1. From Table \ref{missptable}, it is observed that optimal proportions are slightly sensitive to the choice of priors. For example see the optimal allocations corresponding to Uniform(0, 0.2) and Beta(5, 5) priors.  Overall, we may conclude that there is not much change in the results when we use a prior for $\alpha$ instead of some fixed values.

\begin{table}[H]
\caption{Optimal allocation of $D^B$ under different priors for $\alpha$ and efficiency values of $D_{I}$ (Section 7).}
\begin{center}
\begin{tabular}{c|cccccccccc}
 \mbox{}& \multicolumn{4}{c}{$p_{\omega}$} & \mbox{}& \\
 \cline{2-5}
Prior  & AA & AB & BA & BB & Efficiency\\
\hline
Uni(0, 0.2) &0.1520    &0.2700   & 0.2133  &  0.3647  & 0.988\\
Uni(0, 0.5) & 0.1506   & 0.2716   & 0.2161  &  0.3617 & 0.988\\
Uni(0, 0.8) & 0.1503    &0.2744   & 0.2167   & 0.3586 & 0.988\\
Uni(0, 1.0) & 0.1515    &0.2766   & 0.2124   & 0.3595 & 0.988\\
Beta(2, 38)    & 0.2000    &0.2000    &0.3000   & 0.3000 & 0.996\\
Beta(4, 12)   &  0.1997    &0.1992    &0.3005    &0.3006& 0.995\\
Beta(6, 10)    &  0.2060    &0.1986    &0.2908    &0.3047& 0.991\\
Beta(5, 5)   & 0.1843    &0.1841    &0.2876    &0.3440& 0.989\\
\hline
\end{tabular}\label{missptable}
\end{center}
\end{table}

\section{Computer programs used to obtain the optimal designs and estimate the parameters}

Approximation of the multidimensional integrals of the objective functions in equations (\ref{objectivefn}) and (\ref{objectivefnal}) to obtain the optimal designs  is done with Latin Hypercube Sampling (LHS). For uniform priors, we use the average of (\ref{objectivefn}) and (\ref{objectivefnal}) across 100-point discrete samples using LHS as the approximate solution of (\ref{objectivefn}) and (\ref{objectivefnal}), respectively. When $\thetav$ has a Gaussian distribution, Latin Hypercube Sampling from Gaussian fields is used (for more details see \cite{Stein1987}). A MATLAB function \textit{lhsdesign} is used to sample points from the parameter space.
To obtain the optimal  proportions of subjects ($p_{\omega}$) assigned to treatment ($\omega$), \textit{fmincon} function in MATLAB is used. The {\it fmincon} algorithm finds a minimum of a constrained nonlinear multivariable function, and by
default is based on the Sequential Quadratic Programming algorithm. For more details please see the link http://in.mathworks.com/help/optim/ug/fmincon.html$\#$description. A genetic algorithm (\textit{GA} function in MATLAB) verifies the results obtained from the \textit{fmincon}. The estimation  and the estimated confidence intervals of model parameters are done using \textit{GENMOD} procedure in  SAS (\cite{SAS1999}).

{\it Note}: All MATLAB and SAS programs are available to the readers upon request from the first author of this article.

\section{Concluding Remarks}
Crossover designs are popular as designs of choice in many clinical
and pharmaceutical trials for comparing treatments.  However, very often  in these situations the response does not follow the usual assumptions of normality, and  generalized linear models have to be used to model the data.
In this article, we address the designing of such crossover trials when a GLM is fitted. Since the designs are dependent on the model parameters, Bayesian designs are proposed. Comparing our main results based on GLMs with those of normal response models, we see that they are quite similar in many cases.

The main results on
the estimation of direct treatment effects using the proposed $D_A$-optimal Bayesian designs  ($D^B$) are summarized below.
\begin{itemize}
\item{\textbf{For $t=p=4$ when the response is binary :}} Williams design is as efficient as $D^B$ and is seen to perform the best under both CS and AR(1) correlation structures for a reduced as well as a full model.
\item {\textbf{For $p=t=2$ when the response is Poisson distributed:}}
 Design \{$AB, BA$\} has the highest efficiency in a reduced model framework while for a full model, design \{$AB, BA, AA, BB$\} is most efficient. Both designs have equivalent efficiency as $D^B$ for the respective models.
\item {\textbf{For $p=3, t=2$ when the response is Gamma distributed:}}
Under log link function, $D_A$-optimal Bayesian designs are same as in case of normal responses.

For reciprocal link function, under reduced model, design $D_{b}$ (treatment sequences $\{ABB$, $BAA\}$ with equal proportions), perform as well as $D^B$ under the CS correlation structure, while for AR(1) correlation structure, design $D_{c}$ (treatment sequences $\{ABA, BAB,$ $ ABB, BAA\}$ with equal proportions) has the equal efficiency as $D^{B}$.

In case of full model, design $D_{a}$ (treatment sequences $\{ABB,BAA,AAB,BBA\}$ with equal proportions) is equally efficient as $D^{B}$, and performs better than $D_{b}$ and $D_{c}$.
\end{itemize}
In many biological experiments while studying the effect of drugs, the response measured may not be binary in nature but say ordinal. As  an example consider a  $3\times 3$ crossover trial  (cited by  \cite{Jones_2014}) where the effect of three treatments on the amount of patient relief is studied. The response obtained is categorized as none, moderate or complete, making it  ordinal in nature with three categories. Thus, there is a need to address optimal crossover deigns not just for binary models but also for multi categorical responses. In these cases, instead of the logit link, a generalized logit or a proportional odds model may be  used. Also, other than the correlation between measurements from the same subject we would have to consider the relation between response categories.  \cite{Jones_2014} discusses modeling of ordinal data using the GEE approach. In future, we are interested to study D-optimal Bayesian designs for such multicategorical models.

\section*{Appendix}
Consider a finitely supported approximate crossover design with $k$ treatment sequences. The design can be expressed in the form of a probability measure as follows:
\begin{equation*}
\zeta = \begin{Bmatrix}
 \omega_{1}       & \omega_{2} &  \dots & \omega_{k} \\
 p_{\omega_{1}}       &  p_{\omega_{2}} &  \dots &  p_{\omega_{k}}
\end{Bmatrix},
\end{equation*}
where $\omega_{i} \in \Omega$ (set of all treatment sequences considered) and $p_{\omega_{i}}$ is the proportion of subjects assign to treatment sequence $\omega_{i}$ such that $p_{\omega_{i}}\geq 0$ and $\sum_{i = 1}^{k}p_{\omega_{i}} = 1$, for $i = 1, \cdots, k$. Let $M = M(\zeta, \thetav)$ denotes the asymptotic  information matrix of estimates of the parameter vector $\thetav$. This in turn is the reciprocal of the variance-covariance matrix defined in equation (\ref{infmat}). If the interest is in the estimation of a linear combination of the parameters of the form $\lambda = W'\thetav$, where $W$ is a $m\times s$ matrix with rank $s\leq m$. The information matrix of $\lambda$ for a design $\zeta$ is given by $C = C(\zeta, \lambda) = (W'M^{-1}W)^{-1}$. Next theorem insures the optimality of designs obtain for the estimation of $\lambda$ under the prior distribution of $\thetav$.
\begin{theorem}\label{thm1}
Under the GEE model considered for the linear predictor, link function and working correlation, the following conditions for a continuous  design $\zeta^{*}$ are equivalent:

\begin{itemize}

\item[1.] $\zeta^{*}$ minimizes  $\Psi(\mathfrak{B},\zeta,\alpha)$ defined in equation (\ref{objectivefn}), $\forall$\; $\zeta \in \chi$,
where $\chi$ is the set of all possible designs.

\item[2.] $\zeta^{*}$ satisfies the following condition:

\begin{equation}\label{eqv2}
E_{F}\left[\mbox{tr}\left(M(\zeta^{*}, \thetav)^{-1}WCW'M(\zeta^{*}, \thetav)^{-1}\right)M(\zeta_{\omega}, \thetav)\right]\leq s\;  \forall\; \omega \in \Omega,
\end{equation}
where $F$ is the prior distribution of $\thetav$ and $M(\zeta_{\omega}, \thetav)$ is the information matrix with respect to the design $\zeta_{\omega}$ having unit mass at single treatment sequence $\omega$. Equality in equation (\ref{eqv2}) is achieved if any $\omega$ in the Bayesian $D_A$-optimal design is inserted.
\end{itemize}
\end{theorem}

Proof of this theorem follows directly from Theorem 3.1 of \cite{Pettersson2005}. These optimal designs are known as average $D_A$-optimal designs. A similar equivalence theorem is proved and used by \cite{Woods2011} to show the optimality of blocked designs with non-normal responses. Expressions of matrix $W$, $\thetav$ and respective ranks for examples used in this article are:
\begin{itemize}

\item Example 1

\begin{equation*}
W' = \begin{bmatrix}
0&0&0&0&1&0&0&0&0&0\\
0&0&0&0&0&1&0&0&0&0\\
0&0&0&0&0&0&1&0&0&0
 \end{bmatrix}
\end{equation*}
$\thetav = (\nu, \beta_{1}^{*}, \beta_{2}^{*}, \beta_{3}^{*}, \tau_{1}^{*}, \tau_{2}^{*}, \tau_{3}^{*}, \gamma_{1}^{*}, \gamma_{2}^{*}, \gamma_{3}^{*})'$ and $s$ = 3.
\item Example 2
\begin{equation*}
W' = \begin{bmatrix}
0&0&1&0
 \end{bmatrix}
\end{equation*}
$\thetav = (\nu, \beta^{*}, \tau^{*}, \gamma^{*})'$ and $s$ = 1.

\item
Example 3
\begin{equation*}
W' = \begin{bmatrix}
0&0&0&1&0
 \end{bmatrix}
\end{equation*}
$\thetav = (\nu, \beta_{1}^{*}, \beta_{2}^{*}, \tau^{*}, \gamma^{*})'$ and $s$ = 1.

\end{itemize}

\textbf{Acknowledgement:} Mr. Satya Prakash Singh wishes to thank University Grant Commission (UGC), India, for the award of a research fellowship. The work of Siuli Mukhopadhyay
was supported by the IRCC seed funding under the health care consortium [Grant
Number: 15IRSGHC004]. These supports are gratefully acknowledged.
\newpage
\section*{References}
\bibliographystyle{elsarticle-harv}
\bibliography{bayes1}

\end{document}